\documentclass{elsart}
\usepackage[dvips]{graphicx}
\usepackage{psfrag}
\usepackage{subfigure}
\usepackage{flafter}
\begin{document}
\begin{frontmatter}
\title{Microcanonical solution of lattice models with long range interactions}

\author{Julien Barr\'e\thanksref{julien}}

\address{Laboratoire de Physique, UMR-CNRS 5672,
ENS Lyon, 46 All\'{e}e d'Italie, 69364 Lyon C\'{e}dex 07, France}
\address{Dipartimento di Energetica ``S. Stecco", 
Universit\'a di Firenze, Via S. Marta, 
3 I-50139, Firenze, Italy, INFM and INFN, Firenze}

\thanks[julien]{E-mail: Julien.Barre@ens-lyon.fr}

\begin{abstract}
We present a general method to obtain the microcanonical solution of lattice 
models with long range interactions. As an example, we apply it to the long 
range Ising 
chain, focusing on the role of boundary conditions.  
\end{abstract}
\end{frontmatter}
\vspace{-1.25truecm}
\section{Introduction}
\label{Introduction}
Throughout the paper, systems in $d$ dimensions with a pairwise 
interaction potential
which decays at large distances as $V(r)\sim 1/r^{d+\sigma}$ with
$-d\leq \sigma \leq 0$, will be referred to as
systems with {\it long range} interactions \footnote{Such interactions are 
sometimes 
called in the literature {\it nonintegrable}}. Recently, such systems have 
attracted much attention, since they are believed to be in some cases 
candidates for the application of Non Extensive Thermostatistics, introduced 
by Tsallis. Actually, even the standard statistical mechanics 
of these systems are still not well studied.\\ 
Last year, Campa et al.
 \cite{cgm}, 
and, in a more general fashion, Vollmayr-Lee and Luijten 
\cite{Vollmayr}, 
gave the solution within the standard canonical ensemble;      
however, it is known, from the study of self-gravitating systems and other 
models, that long range interactions may produce inequivalences between the 
results of the canonical and microcanonical ensembles 
\cite{pad,lb1,lb2,bmr}. A complete understanding of these models 
requires thus a 
microcanonical solution. We present in this letter a simple 
procedure to obtain the microcanonical solution of long-range interacting 
lattice systems, and illustrate it on the Ising case, focusing on the 
influence of boundary conditions.\\
   
\section{The $\alpha$-Ising model}
To study the influence of the long range interactions, we introduce a 
generalization of the one dimensional Ising model, with an additional 
parameter $\alpha$ controlling the decay of the interaction between two sites.
The hamiltonian reads:\\
\begin{equation}
H =  -\frac{1}{2\tilde{N}}\sum_{i,j}\frac{S_iS_j}{d_{ij}^{\alpha}}
\end{equation}
In this expression, the sum extends over all pairs of sites; $d_{ij}$ is the 
distance between the sites $i$ and $j$, and it depends on the 
boundary 
conditions: if we use periodic boundary conditions, the system must be seen as 
a closed ring, and $d_{ij}=\min (|i-j|,~N-|i-j|)$; if we use free boundary 
conditions, then $d_{ij}=|i-j|$. Finally $\tilde{N}\propto N^{1-\alpha}$ 
is a rescaling factor chosen in order to obtain an extensive energy 
\cite{cgm}; $\alpha=0$ (non decreasing interactions) leads to 
$\tilde{N}=N$, which is the usual rescaling factor for mean-field 
hamiltonians.

\section{The coarse graining procedure}
\begin{figure}
\centering
\psfrag{eigenvalue index}[Bl][Bl][0.5][0]{Eigenvalue index $k/N$}
\psfrag{eigenvalue}[Bl][Bl][0.5][0]{Eigenvalue}
\psfrag{0}[][][.5][0]{$0$}
\psfrag{1}[][][.5][0]{$1$}
\psfrag{0.5}[][][.5][0]{$0.5$}
\psfrag{0.1}{}
\psfrag{0.2}{}
\psfrag{0.3}{}
\psfrag{0.4}{}
\psfrag{0.6}{}
\psfrag{0.7}{}
\psfrag{0.8}{}
\psfrag{0.9}{}
\psfrag{1.2}{}
\psfrag{1.4}{}
\resizebox{8cm}{5cm}{\includegraphics[scale=.4]{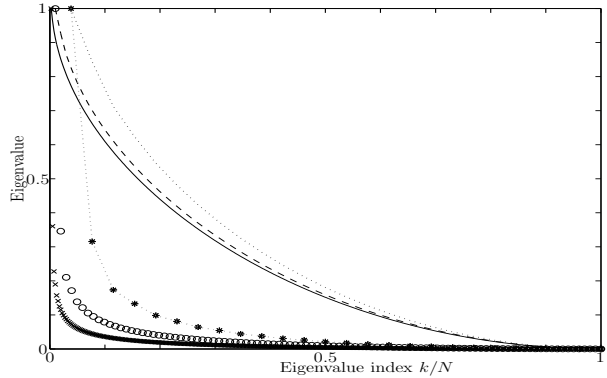}}
\caption{Spectrum of the interaction matrix with periodic boundary 
conditions. The curves correspond to $\alpha=1.5$, for $N=50$ (dotted), 
$N=200$ (dashed), $N=800$ (full line). The symbols correspond to 
$\alpha=0.5$, for $N=50$ (stars), $N=200$ (circles), $N=800$ (crosses).
In the short range case ($\alpha=1.5$) the spectrum tends to a smooth 
curve, whereas in the long range case ($\alpha=0.5$), it shrinks on 
the $y$-axis as $N$ increases, indicating that 
all modes with $k/N\neq 0$ are irrelevant in the thermodynamic limit. }
\label{spectrum}
\end{figure}

To describe the system, we use a coarse grained magnetization 
function $m(x)$,  and we rescale the total length of the 
lattice to one, so that $0\leq x \leq 1$. For a finite chain, with 
a grain size $n$, $m$ would be defined as follows
\begin{equation}
m(jn/N) = \frac{1}{n}\sum_{i=nj+1}^{n(j+1)}S_i\nonumber
\end{equation}
To justify this approach, let us consider the interaction matrix $K_{ij}=
\frac{1}{2\tilde{N}}\frac{1}{d_{ij}^{\alpha}}$. If the boundary conditions 
are periodic, $K$ appears to be cyclic, and is 
explicitly diagonable. One obtains for the spectrum and the eigenvectors (each
eigenvalue corresponds to a two dimensional eigenspace except the $k=0$ one): 
\begin{equation}
 \lambda_k~  =  \frac{2}{\tilde{N}}\sum_{i=1}^{N/2-1}\frac{\cos(2\pi k i/N)}
{i^{\alpha}}
\end{equation}
\begin{displaymath}
 u^{(k)}_j  =  \cos(2\pi k j/N)  
\end{displaymath} 
\begin{displaymath}
v^{(k)}_j  =  \sin(2\pi k j/N) 
\end{displaymath}
This spectrum has a very interesting property. On the one hand, if one fixes 
the wavenumber $k$, the corresponding eigenvalue goes to some finite value 
when $N$ goes to infinity, thanks to the choice of $\tilde{N}$. On the other 
hand, if one fixes $k/N=x$, which amounts to fix the wavelength of the 
eigenmode considered in units of the lattice spacing, the corresponding 
eigenvalue $\lambda(x)$ goes to zero when $N$ goes to infinity, whatever $x$ 
is, provided it is non zero Fig.~\ref{spectrum}. This means that all 
these modes involving a 
possibly large but finite number of spins in one wavelength are irrelevant in 
the $N \to \infty$ limit. Consequently, the coarse graining procedure, which 
drops precisely these short wavelengths modes is adequate to describe the 
system. It is important to notice that under more general boundary 
conditions, the spectrum is not explicitely computable, but heuristic 
arguments and numerical checks show that the qualitative 
properties described above are unchanged.

 Using the coarse grained 
magnetization, we may now write the energy as
\begin{equation}
\label{energy1}
H = -\frac{N}{2} \int_{0}^{1}\int_{0}^{1}  m(x) m(y)
K_{\alpha}(x,y)~dx~dy
\end{equation}
with $K_{\alpha}(x,y)=cste/d(x,y)^{\alpha}$, where the constant is chosen 
through 
$\tilde{N}$ so that the maximal eigenvalue of $K_{\alpha}$ is $1$ (this 
facilitates 
the comparison between different values of $\alpha$). The energy may also be 
written as
$H = -N \sum_k \lambda_k \tilde{m}_k^2$
where $\tilde{m}_k$ is the amplitude of the eigenmode $k$ in the 
function $m(x)$.\\

\section{Expression of the entropy}
Given a magnetization profile $m(x)$, it is possible to determine 
approximately the corresponding number of microscopic configurations, 
using again the fact that $m$ varies slowly, on scales involving an 
infinite number of spins. After some easy combinatorial algebra and the use 
of Stirling formula, the logarithm of the number of configuration reads
\begin{equation}
\label{entropy}
S(m(x)) = -\frac{N}{2} \int_0^1\left[(1+m)~\ln\left(\frac{1+m}{2}\right)~
+(1-m)~\ln\left(\frac{1-m}{2}\right)\right]~dx
\end{equation}
The above expression of the entropy is valid for the Ising model; in general, 
the appropriate form of the entropy has to be derived, through some usually 
easy combinatorial calculus.
To obtain the most probable
magnetization profile in the microcanonical ensemble, we now have to maximize 
$S$ with respect to $m$, keeping $H$ constant\footnote{Let us note that the 
canonical solution is also available, through the minimization of 
$\beta H-S$ at fixed $\beta$}.
\section{Application to the Ising model}
Let's now calculate the most probable magnetization profile for the 
$\alpha$-Ising model using the procedure described above, first for 
periodic boundary conditions, then for free boundary conditions.\\
Using a Lagrange multiplier $\beta$, we can write a necessary and sufficient 
condition to get a stationary point of equation (\ref{entropy}) with the 
constraint given by equation (\ref{energy1}); after a few algebra, we are 
left with
\begin{equation}
\label{statcond}
m(x)=\tanh \left(\beta\int_0^1 K_{\alpha}(x,y)m(y)~dy\right)
\end{equation}
with $\beta$ given by the energy constraint.
If we use periodic boundary conditions, it is easy to see that 
$\int K_{\alpha}(x,y)~dy$ is independent of $x$, so that equation 
(\ref{statcond}) together with the energy constraint
admits a uniform solution for $m(x)$. Furthermore, a second order 
calculation would show that this uniform solution is always a local entropy 
maximum (in other words, the eigenvalues associated to non uniform modes 
are too weak to 
destabilize the uniform solution). On the contrary, if the boundary 
conditions are free, the constants are not eigenvectors of the kernel 
$K_{\alpha}(x,y)$ anymore, and the solutions will be non uniform.\\
To verify these 
predictions, we use a numerical optimization algorithm. Starting from a 
variety of initial magnetization profiles, we always get the same equilibrium 
solution, suggesting that for both type of boundary conditions, there is only 
one entropy maximum.

\begin{figure}
\centering
\psfrag{position on the lattice}[Bl][Bl][0.5][0]{Position on the lattice}
\psfrag{magnetization}[Bl][Bl][0.5][0]{Magnetization}
\psfrag{mag profile for different alpha value}{}
\psfrag{mag profile for periodic boundary conditions}{}
\psfrag{0}[][][.5][0]{$0$}
\psfrag{1}[][][.5][0]{$1$}
\psfrag{0.5}[][][.5][0]{$0.5$}
\psfrag{0.1}{}
\psfrag{0.2}{}
\psfrag{0.3}{}
\psfrag{0.4}{}
\psfrag{0.6}{}
\psfrag{0.7}{}
\psfrag{0.8}{}
\psfrag{0.9}{}
\includegraphics[scale=.4]{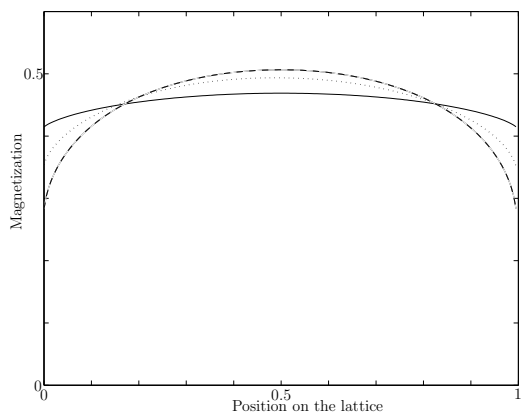}
\caption{ Magnetization profile at energy $-0.1$, for 
free boundary conditions and $\alpha=0.2$ (full line), $\alpha=0.5$ (dotted 
line), $\alpha=0.8$ (dashed line): the system is inhomogeneous. For periodic 
boundary conditions, the magnetization profile is homogeneous and invariant 
under a change of $\alpha$.}
\label{profile}
\end{figure}
As expected, the profile is uniform and independent of $\alpha$ for periodic 
boundary conditions; this is the microcanonical 
counterpart of the universality found in~\cite{cgm}.
Figure \ref{profile} shows the results of the numerical computations for
 free boundary 
conditions: the profile depends on $\alpha$, and becomes more and more 
uniform as $\alpha$ goes to zero, as should be expected.
The $\alpha$-Ising 
model does not display any phase transition in the microcanonical ensemble, 
just as the mean-field one. To study the phase transitions with long range 
interactions, one should introduce a slightly more complex model, for 
instance the Blume-Emery-Griffiths one, \cite{bmr}; work along this line 
is in progress.
 
\section{Conclusion}

We have described a general method to obtain the microcanonical solution of 
lattice models with long range interaction, taking advantage of the quasi 
mean-field structure induced by the long range interactions. Although the 
procedure has been illustrated with the Ising chain, it is straightforward 
to generalize it to various lattice models (Potts models, 
X-Y type models, Blume-Emery-Griffiths model...) in any dimension, by
modifying the expression of the entropy; it will be applied in 
subsequent work to systems displaying phase transitions.\\
The problem reduces to an optimization problem under constraint, which 
is very similar to that obtained after a statistical 
analysis of the two dimensional Euler equation in fluid mechanics   
\cite{robsom}, and where the boundary conditions play a central role. 
Despite the huge gain in complexity with respect to the initial states 
counting problem, it is not easy to solve and requires in some cases 
a numerical treatment.     
\begin{ack}
I warmly thank T.~Dauxois for many suggestions and careful reading of the 
manuscript, and I acknowledge enlightening discussions with S.~Ruffo, 
D.~Mukamel, P.H.~Chavanis. I am grateful to the NEXT 2001 conference for a 
fellowship.
\end{ack}


\begin{thebibliography}{99}
 
\bibitem{cgm} A.~Campa, A.~Giansanti and D.~Moroni , to appear 
in \emph{Chaos, Solitons and Fractals}.
\bibitem{Vollmayr}B.~P.~Vollmayr-Lee and E.~Luijten , \emph{Phys. Rev. E} 
{\bf 63}, 031108 (2001).
\bibitem{pad} T.~Padmanabhan, \emph{Phys. Rep.} {\bf 188}, 285 (1990).
\bibitem{lb1} D.~Lynden-Bell and R. Wood, \emph{Mon. Not. R. Astron. Soc.}
 {\bf 138}, 495 (1968).
\bibitem{lb2} D.~Lynden-Bell, \emph{Physica A} {\bf 263}, 293 (1999).
\bibitem{bmr} J.~Barr\'e, D.~Mukamel and S.~Ruffo,  
in \emph{Phys.~Rev.~Lett.}, {\bf 87}, 030601  (2001).
\bibitem{robsom} R.~Robert and J.~Sommeria, \emph{J. Fluid. Mech} {\bf 229}, 
291 (1991).



\end{thebibliography}
\end{document}